\documentclass[10pt,aps,prl,twocolumn,superscriptaddress,notitlepage,nofootinbib]{revtex4-1}
\usepackage{amsmath}
\usepackage{bm}
\usepackage{epsfig}
\usepackage{color,soul}
\usepackage{multirow}
\def\bea#1\eea{\begin{align}#1\end{align}}

\newcommand{\bef}{\begin{figure}[h!tb]\centering}
\newcommand{\eef}{\end{figure}}

\newcommand{\tvec}[1]{\boldsymbol{#1}}

\allowdisplaybreaks

\begin{document}
\title{A transverse momentum dependent framework for 
\\
back-to-back photon+jet production}

\author{Maarten G.A. Buffing}
\email{mbuffing@physics.ucla.edu}
\affiliation{Department of Physics and Astronomy, University of California, Los Angeles, California 90095, USA}

\author{Zhong-Bo Kang}
\email{zkang@physics.ucla.edu}
\affiliation{Department of Physics and Astronomy, University of California, Los Angeles, California 90095, USA}
\affiliation{Mani L. Bhaumik Institute for Theoretical Physics, University of California, Los Angeles, California 90095, USA}
\affiliation{Theoretical Division, Los Alamos National Laboratory, Los Alamos, New Mexico 87545, USA}

\author{Kyle Lee}
\email{kunsu.lee@stonybrook.edu}
\affiliation{C.N. Yang Institute for Theoretical Physics, Stony Brook University, Stony Brook, NY 11794, USA}
\affiliation{Department of Physics and Astronomy, Stony Brook University, Stony Brook, NY 11794, USA} 

\author{Xiaohui Liu}
\email{xiliu@bnu.edu.cn}
\affiliation{Center of Advanced Quantum Studies, Department of Physics, Beijing Normal University, Beijing 100875, China}

\date{\today}

\begin{abstract}
We propose the $pp \rightarrow \text{jet} + \gamma $ as a new process for studying the Transverse Momentum Dependent Parton Distribution Functions (TMDs). To do so, we developed a novel framework for the jet-$\gamma$ imbalance in $pp \rightarrow \text{jet} + \gamma $ using Soft Collinear Effective Theory. The new framework opens up many new insights to the TMDs which the current TMD studies confined to the Drell-Yan and semi-inclusive deep inelastic scattering processes cannot achieve. The established formalism provides, for the first time, a first principle prediction of the jet-$\gamma$ imbalance in $pp$ collisions and therefore a first direct probe of the factorization breaking effects when compared with the future experimental data. If the factorization breaking effects are found small, the process will offer the unique sensitivities to both the polarized and un-polarized quark/gluon TMDs. We demonstrate the predictive power of the framework by calculating each component in the formalism to the next-to-leading order accuracy and by realizing the next-to-leading-logarithmic evolution. We provide the first numerical estimates for this process, for both the unpolarized cross section and the Sivers asymmetry.
\end{abstract}

\date{\today}

\maketitle

{\it Introduction.} The theoretical study and experimental exploration of the internal structure of nucleons are of fundamental importance to science and have recently entered a new exciting phase~\cite{Boer:2011fh,Accardi:2012qut}. In the past decades an understanding of nucleons in terms of quarks and gluons (partons), the degrees of freedom of Quantum Chromodynamics (QCD), has successfully emerged. Progress has been made in constructing a ``one-dimensional'' picture of the nucleon~\cite{Gao:2017yyd}, in the sense that we ``only'' know about the longitudinal motion of partons in fast moving nucleons. In the last few years, theoretical breakthroughs have paved the way to extending this simple picture to the transverse as well as longitudinal momentum space, i.e. three dimensions. This new information is encoded in the novel concept of ``Transverse Momentum Dependent Parton Distribution Functions'' (TMDs), which help address the long-standing questions concerning the confined motion of quarks and gluons inside the nucleon~\cite{Collins:2011zzd}. Besides providing information about how the partons move in the transverse plane, they also probe a variety of spin-spin and spin-momentum correlations when the parton and nucleon spin states are considered~\cite{Boer:2011fh,Accardi:2012qut,Aschenauer:2015eha}.

Extractions of TMDs from the experimental data through a global analysis have become extremely active in recent years. For example, the unpolarized quark TMDs have been extracted by several groups~\cite{Bacchetta:2017gcc,Scimemi:2017etj,Su:2014wpa,Anselmino:2013lza}, which represent the distributions of unpolarized quarks inside an unpolarized nucleon. Beyond the unpolarized TMDs, the most interesting and studied polarized TMD is probably the Sivers functions~\cite{Sivers:1989cc,Sivers:1990fh}. The Sivers functions were first thought to be vanishing \cite{Collins:1992kk}, but later shown to be nonzero~\cite{Brodsky:2002cx,Collins:2002kn,Brodsky:2002rv,Boer:2003cm}, corresponding to unpolarized quarks in a transversely polarized proton. For recent extraction of the quark Sivers functions, see Refs.~\cite{Echevarria:2014xaa,Boglione:2018dqd,Sun:2013hua}.

To extract these TMDs, QCD factorization formalisms have to be established and proven, which express the cross sections as a convolution of partonic hard functions and the associated non-perturbative TMDs. So far, only two processes have well-established TMD factorization frameworks from which the Sivers functions have been or are being probed: semi-inclusive deep inelastic lepton-hadron scattering (SIDIS) and Drell-Yan process in proton-proton collisions~\cite{Collins:2011zzd}. Remarkably, the Sivers function is predicted to have opposite signs in these two processes~\cite{Brodsky:2002cx,Collins:2002kn,Brodsky:2002rv,Boer:2003cm}, which has been explored in both COMPASS~\cite{Aghasyan:2017jop} and RHIC~\cite{Adamczyk:2015gyk} experiments. Thus far, these two processes are the only ones in which the Sivers functions are being studied. In order to expand the possibilities for probing the Sivers functions and TMDs in general, it would, therefore, be extremely useful to have access to more processes. 

In this letter, to explore along this direction, we study back-to-back photon-jet production in both unpolarized and transversely polarized proton-proton collisions, $p(p_1, s_\perp) + p(p_2) \rightarrow \text{jet} + \gamma$, where $p_1(p_2)$ are the momenta of the incoming protons, and $s_\perp$ is the transverse polarization of the proton. Such a process was proposed already in~\cite{Bacchetta:2007sz} and was studied within a leading-order parton model formula. However, later theoretical studies found factorization breaking effects for such processes~\cite{Collins:2007nk,Rogers:2010dm} and thus further investigations have been hindered. Recently there has been a great interest in studying such a process in experiments~\cite{Adare:2016bug,Aidala:2018bjf} to probe any factorization breaking. Because of that, it becomes urgent to develop a theoretical framework so that the experimental measurements can be compared to assess any potential factorization breaking. In this letter we will precisely achieve this goal. We propose a framework for studying TMDs in this process within Soft Collinear Effective Theory (SCET)~\cite{Bauer:2000ew,Bauer:2000yr,Bauer:2001ct,Bauer:2001yt,Bauer:2002nz}. See other work along this line~\cite{Zhu:2012ts,Sun:2015doa,Sun:2018icb}. We study both unpolarized differential cross section and Sivers asymmetry, and make a numerical estimate for the relevant RHIC kinematics. 

{\it TMD formalism.} 
We study jet-photon production in proton-proton collisions, $p(p_1, s_\perp) + p(p_2)\rightarrow \text{jet} + \gamma$, where $s=(p_1+p_2)^2$, and the transverse momenta of the jet and the photon are given by $p_{J\perp}$ and $p_{\gamma\perp}$, respectively. Here the transverse momentum is measured with respect to the beam axis in the center-of-mass frame of the colliding protons. One defines the imbalance $\vec{q}_\perp$ between the transverse momenta of the photon and the jet, and the average of the transverse momenta $p_\perp$ as
\bea
\vec{q}_\perp \equiv \vec{p}_{\gamma\perp}+ \vec{p}_{J\perp}, 
\qquad
p_\perp = \left|\vec{p}_{\gamma\perp} - \vec{p}_{J\perp}\right|/2. 
\eea
The rapidities of the jet and the photon are given by $y_J$ and $y_\gamma$, respectively. We focus on the back-to-back region, i.e. the kinematic regime where the imbalance is much smaller than the average transverse momentum: $q_\perp \equiv |\vec{q}_\perp| \ll p_\perp$. In this region, within SCET framework where Glauber modes are not considered~\cite{SCETG}, one can write down a factorized form for the unpolarized differential cross section as
\begin{align}
\frac{d\sigma}{d{\cal PS}} = & 
\sum_{a,b,c} \int d \phi_J \, 
\int \prod_i^4 d^2\vec{k}_{i\perp}  \delta^{(2)}(\vec{q}_\perp - \sum_i^4 \vec{k}_{i\perp})
\nonumber \\    
 &
  \times f_a^{\rm unsub}(x_a,k_{1\perp}^2)    f_b^{\rm unsub}(x_b,k_{2\perp}^2) S^{\text{global}}_{n{\bar n} n_J }(\vec{k}_{3\perp}) 
 \nonumber \\
 &
  \times   S^{cs}_{n_J}(\vec{k}_{4\perp},R) 
  H_{a b\rightarrow c \gamma}(p_\perp)   J_{c}(p_\perp R) \,,
\label{e:PJ_factorization} 
\end{align}
where the phase space $d{\cal PS} = dy_J dy_\gamma dp_\perp d^2 \vec{q}_{\perp}$, and we have integrated over the azimuthal angle of the jet $\phi_J$ and suppressed the scale dependence for convenience. Here $H_{ab\to c\gamma}(p_\perp)$ are the partonic hard functions, with two relevant partonic channels $q\overline q \rightarrow g \gamma$ and $q g\rightarrow q \gamma$, which are known to two loops~\cite{Becher:2009th}. At the same time, $f_a^{\rm unsub}(x_a,k_{1\perp})$ and $f_b^{\rm unsub}(x_b, k_{2\perp})$ are the so-called un-subtracted TMDs of parton flavors $a$ and $b$~\cite{Collins:2011zzd}, with 
$x_a = \frac{p_\perp}{\sqrt{s}}\left(e^{y_J} + e^{y_\gamma}\right),~x_b = \frac{p_\perp}{\sqrt{s}}\left(e^{-y_J} + e^{-y_\gamma}\right)$. $J_c(p_\perp R)$ is the jet function with the jet size parameter $R$, which encodes the parton $c$ initiating energetic collinear radiations inside the jet. 

Finally, we have two soft functions. $S^{\text{global}}_{n{\bar n} n_J }$ is the wide-angle global soft function, which cannot resolve the small jet cone and thus is jet radius $R$-independent. Here $n$, ${\bar n}$ and $n_J$ are unit light-like vectors pointing along the $z$, $-z$ and the jet axis directions, respectively. On the other hand, $S^{cs}_{n_J}(\vec{k}_{4\perp}, R)$ is the collinear-soft function that captures so-called soft-collinear modes, i.e., for the soft radiations with separations of order $R$ along the jet direction and can thus resolve the jet cone~\cite{Chien:2015cka,Hornig:2017pud}. We emphasize that the transverse momentum $\vec{k}_{4\perp}$ in the collinear-soft function $S^{cs}_{n_J}$ is fixed to be aligned with the jet axis up to power corrections of order $R^2$: $\vec{k}_{4\perp} \approx k_{4\perp} \vec{n}_{J\perp}+{\cal O}(R^2)$. Such corrections can be neglected for highly collimated jets, i.e. jets of small $R\ll 1$. Therefore the collinear-soft function $S^{cs}_{n_J}$ only relies on the magnitude of $k_{4\perp}$ in the end.

Following the usual wisdom of TMD framework, we can Fourier transform Eq.~\eqref{e:PJ_factorization} to the coordinate $b$-space to find
\begin{align}
 \frac{d\sigma}{d{\cal PS}} = & \sum_{a,b,c} \int d\phi_J\int \frac{d^2\vec{b}}{(2\pi)^2} e^{i \vec{q}_\perp\cdot \vec{b}} 
 \nonumber \\
& \times
f_a^{\rm unsub}(x_a, b) f_b^{\rm unsub}(x_b, b) {S}^{\text{global}}_{n{\bar n}n_J}(b \,, c_{\phi_{bj}})
 \nonumber \\
& \times  
 S_{n_J}^{cs}(b \, c_{\phi_{bj}},R)  
H_{a b\rightarrow  c\gamma}(p_\perp)  J_{c}(p_\perp R) \,,
\label{b_factorization}
\end{align}
where we introduce the notation $c_{\phi_{bj}} \equiv \cos \phi_{bj}$ where $\phi_{bj}$ is the azimuthal angle difference between two-dimensional $\vec{b}$ and the jet axis $n_J$. Here ${S}^{\text{global}}_{n{\bar n}n_J}(b \,, c_{\phi_{bj}})$ and $S_{n_J}^{cs}(b \, c_{\phi_{bj}},R)$ are the Fourier transformation of the global soft and collinear-soft functions, with both magnitude $b = |\vec{b}|$ and azimuthal angle dependence. It is important to realize that the global soft function in SIDIS and Drell-Yan processes only depends on the magnitude $b$, but not the azimuthal angle, simply because the soft radiations are uniformly distributed in the azimuthal plane. The novel feature for photon+jet production lies in the fact that when there is a final-state jet, the jet momentum sets a specific direction in the transverse plane, which breaks the uniformity of the soft radiations. We thus have to separately evolve ${S}^{\text{global}}_{n{\bar n}n_J}$ and $S_{n_J}^{cs}$ through their corresponding renormalization group equations and then combine them. 

Now we turn to the Sivers asymmetry. Assuming one of the incoming proton beam is transversely polarized with transverse spin $s_\perp$, the factorized form for the spin-dependent cross section $d\Delta \sigma = \left[d\sigma(s_\perp) - d\sigma(-s_\perp)\right]/2$ is given by
\begin{align}
\frac{d\Delta\sigma}{d {\cal PS}} =&  \epsilon^{\alpha\beta}s_\perp^\alpha \sum_{a,b,c} \int d\phi_J \int \prod_i^4 d^2\vec{k}_{i\perp}  \delta^{(2)}(\vec{q}_\perp - \sum_i^4 \vec{k}_{i\perp})
\nonumber \\
& 
\hspace{-7mm} \times \frac{k_{1\perp}^{\beta}}{M}    f_{1T,a}^{\perp\,\rm SIDIS}(x_a, k_{1\perp}^2) f_b^{\rm unsub}(x_b,k_{2\perp}^2) 
\nonumber \\
&
\hspace{-7mm} \times 
S_{n{\bar n}n_J}(\vec{k}_{3\perp})
S^{cs}_{n_J}(\vec{k}_{4\perp},R) 
  H^{\text{Sivers}}_{a b\rightarrow c \gamma}(p_\perp)  J_{c}(p_\perp R) \, ,
 \label{Sivers_factorization} 
\end{align}
where $\epsilon^{\alpha\beta}$ is a two-dimensional transverse epsilon tensor with $\epsilon^{12} = +1$, and $M$ is the proton mass. In general, the Sivers functions are not universal, and are different for different partonic processes. One well-known example is the aforementioned sign change between SIDIS and Drell-Yan processes. For the Sivers functions probed in photon+jet production in transversely polarized proton-proton collisions, they are related to those in SIDIS in a slightly more complicated form. Such process-dependence can be implemented into the hard functions $H^{\text{Sivers}}_{a b\rightarrow c \gamma}$, at least to leading-order. As shown in~\cite{Qiu:2007ey}, $H^{\text{Sivers}}_{a b\rightarrow c \gamma}$ are obtained by multiplying the unpolarized hard function by factors of $\frac{N_c^2+1}{N_c^2-1}$ for $q{\bar q}\to g\gamma$ channels while $-\frac{N_c^2+1}{N_c^2-1}$ for the $qg\to q\gamma$ channel. Once such process-dependences are taken into account within the hard functions $H^{\text{Sivers}}_{a b\rightarrow c \gamma}$, we are left with the same Sivers functions as probed in SIDIS, $f_{1T,a}^{\perp\,\rm SIDIS}(x_a, k_{1\perp}^2)$. Here the superscript ``SIDIS'' emphasizes that it is the same as the Sivers function probed in SIDIS process. At the end of the day, a similar Fourier transformation from transverse momentum space to coordinate $b$-space can be performed for spin-dependent cross section in Eq.~\eqref{Sivers_factorization}.

{\it Relation to the standard TMDs.} We want to establish connections/relations between the above TMDs and the standard TMDs. By standard TMDs, we mean those probed in SIDIS process. To explicitly demonstrate this, we compute all the relevant functions to next-to-leading order (NLO). Some of them are already available in the literature. The NLO jet functions are computed long time ago and can be found in~\cite{Liu:2012sz,Ellis:2010rwa}. The un-subtracted TMDs are available in~\cite{Aybat:2011zv,Echevarria:2016scs,Kang:2017glf}. It is instructive to realize that both the un-subtracted TMDs and the soft functions contain ultraviolet divergences, as well as the so-called rapidity singularities~\cite{Collins:2011zzd}. However, in the final cross section, both divergences cancel among different factors in the factorized formalism, which serves as a consistent check for the formalism. We have explicitly verified this cancelation. 

Let us now turn to the soft functions. We start with the global soft function, whose generic form at NLO in $b$-space takes~\cite{Hornig:2017pud}
\begin{align}
S^{\text{global}}_{n{\bar n}n_J} & = 
1 + \sum_{i<j}\left[\tvec{T}_{i}\cdot\tvec{T}_{j}S_{i j}^{(1)} + {h.c.}\right]\,
\label{e:soft_decomp}
\end{align}
where we have chosen to normalize it to be 1 at LO. With explicit results computed, it is not difficult to note that the global soft function ${S}^{\text{global}}_{n{\bar n}n_J}(b \,, c_{\phi_{bj}})$ can be further factorized as
\bea
{S}^{\text{global}}_{n{\bar n}n_J}(b \,, c_{\phi_{bj}})
= S_{n {\bar n} }(b) S_{n{\bar n} n_J}(b,c_{\phi_{bj}}) \,, 
\label{s_ref}
\eea
where $S_{n {\bar n} }(b)$ only depends on the magnitude $b$, and is exactly the same as the soft function in SIDIS process. It is important and instructive to note that all the rapidity singularities are all contained in $S_{n {\bar n} }(b)$ and are absent within the remnant part $S_{n{\bar n} n_J}(b,c_{\phi j})$. Such a relation is checked through explicit calculations at next-to-leading order (NLO). As the proper TMD is defined through a product of the un-subtracted TMD and the corresponding standard soft function~\cite{Collins:2011zzd}, then once $S_{n {\bar n} }(b)$ is combined with the un-subtracted TMDs $f_a^{\rm unsub}(x_a,b)$ and $f_b^{\rm unsub}(x_b, b)$, then they become standard TMDs as probed in SIDIS, which will be denoted as $f_a(x_a,b)$ and $f_b(x_b, b)$ below. In other words, 
\bea
f_a^{\rm unsub}(x_a,b) f_b^{\rm unsub}(x_b, b) S_{n {\bar n} }(b) = f_a(x_a,b) f_b(x_b, b),
\nonumber
\eea
where there are no rapidity divergences in $f_a(x_a,b)$, nor in $f_b(x_b, b)$ on the right hand side. We can thus use the standard TMDs extracted from SIDIS in the phenomenological studies. 

On the other hand, the remnant part $S_{n{\bar n} n_J}$ at NLO is given by
\begin{align}
S_{n{\bar n}n_J}(b,c_{\phi_{bj}}) =&  1 + \frac{\alpha_s}{\pi}  \tvec{T}^2_c \,
\bigg[
 \ln^2 \left( i \frac{2\mu c_{\phi_{bj}}}{\mu_b}\right)
 + \frac{\pi^2}{8}
-F(c_{\phi_{bj}})
\bigg] \nonumber \\
& - \frac{\alpha_s}{\pi}
(\tvec{T}_b^2 - \tvec{T}_a^2)   \,  y_J \, \ln \left(\frac{\mu^2}{\mu_b^2}\right)
\,,
\end{align}
where $y_J$ is the rapidity of the observed jet, and $\mu_b = 2e^{-\gamma_E}/b$. The color factors are given by $\tvec{T}_{i}^2 = C_F (C_A)$ for $i=q (g)$. Here $F(c_{\phi_{bj}})$ satisfies the condition
$\int_0^{2\pi}\mathrm{d}\phi_{bj} F(c_{\phi_{bj}}) = 0$. At the same time, the NLO collinear-soft function $S_{n_J}^{sc}$ in the $b$-space is found to be
\begin{align}
 S_{n_J}^{(1)}(b,c_{\phi_{bj}}) & = 1
- \frac{\alpha_s \tvec{T}_c^2}{2\pi}
\bigg[
2\ln^2 \left(   i\frac{2\mu c_{\phi_{bj}}}{\mu_b \, R} \right) + \frac{\pi^2}{4}
 \label{e:PJ_S_jet_1}
\bigg]
\,,
\end{align}
The detailed renormalization-group evolved results and the expressions for the final cross section will be given in details in a forthcoming long paper. 

{\it Phenomenology.} 
We will now perform some numerical studies to verify/test our formalism. The photon+jet measurements have been performed at the LHC, see for example Ref.~\cite{Aaboud:2017kff}. The differential cross section is presented as a function of various observables, such as jet or photon transverse momentum, their invariant mass, their relative azimuthal angle, and etc. However, there are no readily available data as a function of transverse momentum imbalance $q_\perp$. Since it has also been shown in~\cite{Aaboud:2017kff} that the default Pythia 8 event generator can describe the dependence on all these observables rather well, we feel that a comparison with Pythia 8 simulation for the $q_\perp$-distribution can also be quite beneficial. Since RHIC would be the only place to perform transversely polarized proton-proton collisions to measure Sivers asymmetry for this process, we make our simulation and comparison for RHIC kinematics. In the numerical calculations, we take the standard unpolarized quark and gluon TMDs from~\cite{Kang:2015msa,Su:2014wpa}.  

In Fig.~\ref{fig:unpolarized}, we plot the normalized differential cross section of back-to-back photon+jet production in unpolarized proton-proton collisions at RHIC energy $\sqrt{s} = 500$ GeV, as a function of transverse momentum imbalance $q_\perp$. We choose the jet transverse momentum $10~\rm{GeV} < p_\perp < 50~\rm{GeV}$ , and integrate over the rapidities for the jet and the photon as, $|y_{\gamma, J}| < 2.0$. The jet is reconstructed via the anti-$k_T$ algorithm with jet radius $R=0.7$. The blue solid histogram is the result from the default Pythia 8 simulation, which has the hadronization turned on. Since our factorized formalism is at the parton level, we also plot the Pythia simulation with hadronization turned off, as shown in the blue dashed histogram. Our theoretical calculation is shown as the red solid curve. As one can clearly see, the numerical results based on our factorized formalism give a very good description of the Pythia simulation. Since the default Pythia 8 turns out to be describing the actual photon+jet LHC data very well, the agreement in Fig.~\ref{fig:unpolarized} is thus very encouraging, which might point to a small factorization breaking in such a process. If this were true, such a process would be very useful for TMD extractions in the future. In particular, since such a process is typically dominated by $qg\to q+\gamma$ and thus would be very sensitive to gluon TMDs. 

\bef
\includegraphics[width=3.0in]{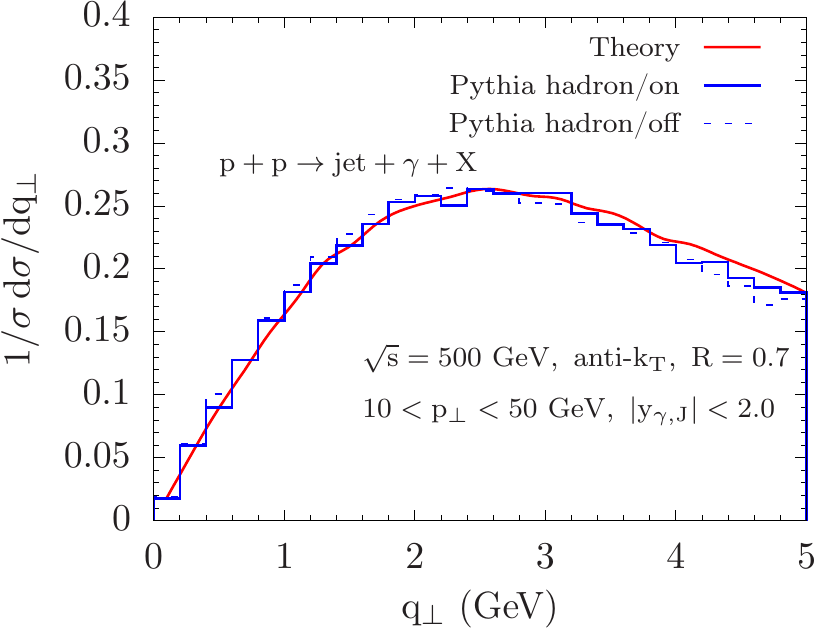} 
\caption{The normalized differential cross section of back-to-back photon+jet production in unpolarized p+p collisions at RHIC energy $\sqrt{s} = 500$ GeV, $p+p\to {\rm jet}+\gamma+X$, as a function of transverse momentum imbalance $q_\perp$.}
\label{fig:unpolarized}
\eef

In Fig.~\ref{fig:sivers}, the Sivers asymmetry $A_N = d\Delta\sigma/d\sigma$ in single transversely polarized proton-proton collisions at RHIC energy $\sqrt{s} = 500$ GeV, is plotted as a function of $q_\perp$. We use the Sivers functions extracted from SIDIS process in~\cite{Echevarria:2014xaa}. We choose to integrate the jet transverse momentum $10~\rm{GeV} < p_\perp < 50~\rm{GeV}$, and rapidities of the photon and the jet are $y_\gamma=y_J = 1$. We find that the asymmetry is on the order of $1\%$ with the TMD evolution as implemented in~\cite{Echevarria:2014xaa}, which should be measurable at RHIC~\cite{Abelev:2007ii}. Since \cite{Echevarria:2014xaa} probably has too strong TMD evolution effect, which is likely to suppress the Sivers asymmetry more than what was observed in the experimental data, the actual Sivers asymmetry for photon+jet production could be much larger.

\bef
\includegraphics[width=3.1in]{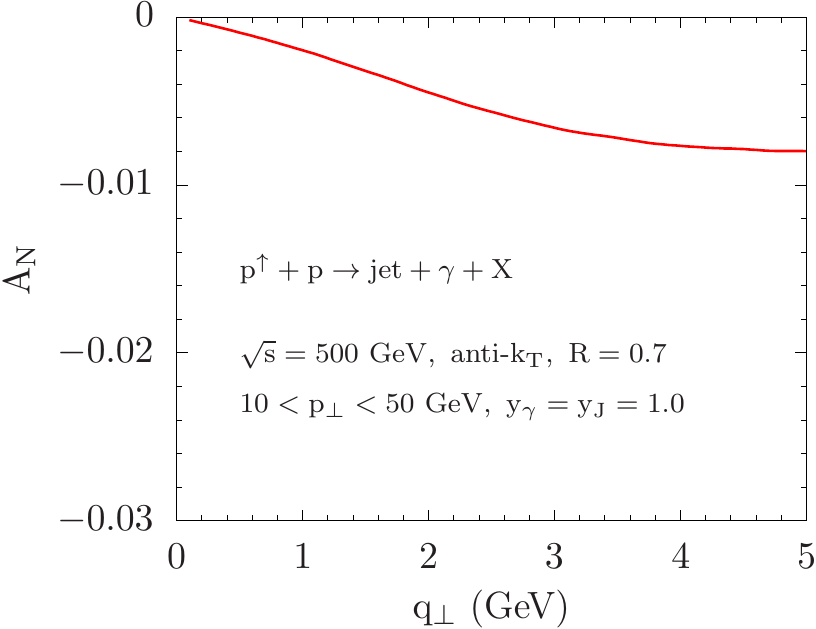} 
\caption{The Sivers asymmetry $A_N$ in single transversely polarized p+p collisions, $p^{\uparrow}+p\to {\rm jet}+\gamma+X$, at RHIC energy $\sqrt{s} = 500$ GeV. }
\label{fig:sivers}
\eef

{\it Conclusions.} In this work, we build up, out of the first principle, a factorized framework for investigating the photon+jet imbalance in proton-proton collisions. The framework shows its sensitivity to both the quark and the gluon TMDs in the unpolarized collisions and to the Sivers functions in the polarized case. 
The framework lays down the theoretical foundation of the future feasible experimental measurements of the process. We demonstrate the predictive power of the formalism by presenting explicitly the NLO results of all ingredients in the factorized formalism and meanwhile, realizing the NLL evolution to resum all the large logarithmic contributions. We further show the first numerical prediction of the imbalance distribution for both the unpolarized cross section and the Sivers asymmetry at the RHIC kinematics. Once the data are available, any deviations of the data from our theory predictions to be observed will shed light on the factorization breaking effects. On the other hand, if the factorization breaking effects are found to be small, the formalism will provide the unique opportunity for studying the gluon TMDs or the Sivers functions by suitably controlling the jet or photon transverse momentum as well as rapidities, which can not be probed in more traditional semi-inclusive deep inelastic scattering and Drell-Yan processes. 

{\it Acknowledgements.} We thank C.~Aidala, E.~C.~Aschenauer, and R.~Seidl for their strong interest and useful discussions on the subject, and Y.~Makris, J.~Qiu, I.~Stewart, and A.~Vladimirov for suggestions and comments. This work is supported by the National Science Foundation under Contract No.~PHY-1720486 (M. B. and Z.K.) and No.~PHY-1620628 (K.L.), by the National Natural Science Foundation of China under Grant No.~11775023 and the Fundamental Research Funds for the Central Universities (X.L.), and by the U.S. Department of Energy, Office of Science, Office of Nuclear Physics under Award No.~DE-AC05-06OR23177, within the framework of the TMD Topical Collaboration. 

\bibliographystyle{h-physrev}
\bibliography{PP_photon-jet}

\end{document}